\documentclass[11pt]{article}

\usepackage{amsmath,amsthm,amssymb,euscript,amscd,graphicx, color, amsfonts, fleqn }

\usepackage{graphicx}

\definecolor{mybackgroundcolor}{rgb}{1,0,1}
\definecolor{plum}{rgb}{.5,0,.5}
\definecolor{dkred}{rgb}{.5,0,0}
\definecolor{ddkred}{rgb}{.35,0,0}
\definecolor{dkblue}{rgb}{.1,0,.6}
\definecolor{ddkblue}{rgb}{0,0,.25}
\definecolor{dkgreen}{rgb}{0,.5,0}
\definecolor{ddkgreen}{rgb}{0,.35,0}
\definecolor{dkgr2}{rgb}{0,.57,0}
\definecolor{dkgr3}{rgb}{0,.64,0}
\definecolor{dkgr4}{rgb}{0,.71,0}

\pagecolor{white}

\newcommand{\be}{\begin{equation}}
\newcommand{\ee}{\end{equation}}
\newcommand{\bd}{\begin{displaymath}}
\newcommand{\ed}{\end{displaymath}}
\newcommand{\bea}{\begin{eqnarray}}
\newcommand{\eea}{\end{eqnarray}}
\newcommand{\rhp}{\mathbb{C}_{+}}
\newcommand{\Hi} {{\cal H}^\infty}

\begin{document}
\title{On the Mixed Sensitivity Minimization for \\ Systems with
Infinitely Many Unstable Modes\thanks{This work was supported in
part by the AFRL/VA and AFOSR under the contract F33615-01-2-3154;
and by the NSF under the grant no.ANI-0073725. }}
\author{ Suat G\"um\"u\c{s}soy\thanks{Dept.~of Electrical and Computer Eng.,
Ohio State University, Columbus, OH 43210, gumussoy.1@osu.edu} \and
Hitay \"Ozbay
\thanks{Dept.~of Electrical and Electronics Eng., Bilkent
University, Bilkent, Ankara TR-06800, Turkey, on leave from
Dept.~of Electrical and Computer Eng., Ohio State University,
Columbus, OH 43210, ozbay.1@osu.edu}}

\date{}
\maketitle

\setcounter{page}{1}

\begin{abstract}

In this note we consider a class of linear time invariant systems with
infinitely many unstable modes. By using the parameterization of all
stabilizing controllers and a data transformation, we show that $\Hi$
controllers for such systems can be computed using the techniques
developed earlier for infinite dimensional plants with finitely many
unstable modes.
\end{abstract}

\section{Introduction}
\setcounter{equation}{0}

It is well known that $\Hi$ controllers for linear time invariant
systems with finitely many unstable modes can be determined by various
methods, see e.g.
\cite{BP02,CZ96,DGS95,FY94,FTZ86,HYT00,KKYYcdc03,OST93,TO95,ZK87}. The
main purpose of this note is to show that $\Hi$ controllers for
systems with infinitely many unstable modes can be obtained by the same
methods, using a simple data transformation. An example of such a
plant is a high gain system with delayed feedback (see Section~3).
Undamped flexible beam models, \cite{Halevi02}, may also be considered
as a system with infinitely many unstable modes.

In earlier studies, e.g. \cite{TO95}, $\Hi$ controllers are computed
for weighted sensitivity minimization involving plants in the form
\be
P(s)=\frac{M_n(s)}{M_d(s)}N_o(s)
\ee
where $M_n(s)$ is inner and infinite dimensional, $M_d(s)$ is inner and
finite dimensional, and $N_o(s)$ is the outer part of the plant, that
is possibly infinite dimensional. In the weighted sensitivity
minimization problem, the optimal controller achieves the minimum
$\Hi$ cost, $\gamma_{opt}$, defined as
\be
\label{eq:wsm} \gamma_{opt}=\inf_{C~{\rm stabilizing}~P}
\left\|
\left[\begin{array}{c}
  W_1(1+PC)^{-1} \\
  W_2PC(1+PC)^{-1}
\end{array} \right]\right\|_\infty ,
\ee
where $W_1$ and $W_2$ are given finite dimensional weights. Note that
in the above formulation, the plant has finitely many unstable modes,
because $M_d(s)$ is finite dimensional, whereas it may have infinitely
many zeros in $M_n(s)$. In this note, by using duality, the mixed
sensitivity minimization problem will be solved for plants with
finitely many right half plane zeros and infinitely many unstable
modes.

In Section~2 we show the link between the two problems and give the
procedure to find optimal $\Hi$ controllers by using the procedure of
the book by Foias {\it et al.}, \cite{FOT}. In Section 3, a delay
system example is given and the design steps for optimal controller are
explained. Concluding remarks are made in Section~4.

\section{Main Result}

Assume that the plant to be controlled has infinitely many unstable
modes, finitely many right half plane zeros and no direct transmission
delay. Then, its transfer function is in the form $P=\frac{N}{M}$,
where $M$ is inner and infinite dimensional (it has infinitely many
zeros in $\rhp$, that are unstable poles of $P$), $N=N_{i} \; N_{o}$
with $N_{i}$ being inner finite dimensional, and $N_o$ is the outer
part of the plant, possibly infinite dimensional. For simplicity of the
presentation we further assume that $N_o,N_{o}^{-1}\in \Hi$.

To use the controller parameterization of Smith, \cite{S89}, we first
solve for $X,Y\in \Hi$ satisfying
\be N X+M Y=1
\label{Bezout} ~~~~~{\rm i.e.}~~~
X(s)=\left(
\frac{1-M(s)Y(s)}{N_{i}(s)} \right) \; N_{o}^{-1}(s).
\ee
Let $z_{1},...,z_{n}$ be the zeros of $N_{i}(s)$ in $\rhp$, and again
for simplicity assume that they are distinct. Then, there are finitely
many interpolation conditions on $Y(s)$ for $X(s)$ to be stable, i.e.
\bd
Y(z_{i})=\frac{1}{M(z_{i})}.
\ed
Thus by Lagrange interpolation, we can find a finite dimensional $Y\in
\Hi$ and infinite dimensional $X \in \Hi$ satisfying (\ref{Bezout}),
and all controllers stabilizing the feedback system formed by the
plant $P$ and the controller $C$ are parameterized as follows,
\cite{S89},
\be
C(s)=\frac{X(s)+M(s)Q(s)}{Y(s)-N(s)Q(s)} \quad \text{where} \; Q(s)
\in \Hi ~ \text{and} \; (Y(s)-N(s)Q(s))\neq 0 .
\ee

Now we use the above parameterization in the sensitivity minimization
problem.  First note that,
\bd
(1+P(s)C(s))^{-1}=M(s)(Y(s)-N(s)Q(s))
\ed
\be
P(s) C(s)(1+P(s)C(s))^{-1}=N(s)(X(s)+M(s)Q(s)).
\ee
Then,
\be
 \label{twobl}
 \inf_{C~{\rm stabilizing}~P}
\left\| \left[\begin{array}{c}
W_{1}(1+PC)^{-1} \\
W_{2} P C(1+PC)^{-1} \end{array} \right] \right\|_{\infty} =
\inf_{Q\in \Hi ~{\rm and}~ Y-NQ\neq 0} \left\| \left[\begin{array}{c}
W_{1}(Y-N Q) \\
W_{2}N(X+M Q) \end{array} \right] \right\|_{\infty}
\ee
where $W_1$ and $W_2$ are given finite dimensional (rational) weights.
From (\ref{Bezout}) equation, we have
\be
 \left\| \left[\begin{array}{c}
W_{1}Y-W_{1}NQ \\
W_{2} N\left( \frac{1-MY}{N} \right) +W_{2}MNQ  \end{array} \right]
 \right\|_{\infty} = \left\| \left[\begin{array}{c}
W_{1}(Y-N_{i}(N_{o}Q)) \\ W_{2}(1-M(Y-N_{i}(N_{o}Q)))  \end{array}
\right] \right\|_{\infty}.
\ee
Thus, the $\Hi$ optimization problem reduces to
\begin{equation} \label{eq:prob1}
\gamma_{opt}=\inf_{Q_{1} \in \Hi ~{\rm and}~Y-N_iQ_1\neq 0}
\left\|
\left[\begin{array}{c} W_{1}(Y-N_{i}Q_{1}) \\
W_{2}(1-M(Y-N_{i}Q_{1}))  \end{array} \right]
\right\|_{\infty}
\end{equation}
where $Q_{1}=N_{o}Q$, and note that $W_{1}(s), W_{2}(s),
N_{i}(s), Y(s)$ are rational functions, and $M(s)$ is inner infinite
dimensional.

The problem defined in (\ref{eq:prob1}) has the same structure as
the problem dealt in Chapter~5 of the book by  Foias, \"Ozbay and
Tannenbaum  (F\"OT)), \cite{FOT} (that is based on \cite{OST93}),
where skew Toeplitz approach has been used for computing $\Hi$
optimal controllers for infinite dimensional systems with finitely
many right half plane poles. Our case is the dual of the problem
solved in \cite{FOT,OST93}, i.e., there are infinitely many poles
in $\rhp$, but the number of zeros in $\rhp$ is finite. Thus by
mapping the variables as shown below, we can use the results of
\cite{FOT,OST93} to solve our problem: \bd
W_{1}^{F\ddot{O}T}(s)=W_{2}(s) \ed \bd
W_{2}^{F\ddot{O}T}(s)=W_{1}(s) \ed \bd X^{F\ddot{O}T}(s)=Y(s) \ed
\bd Y^{F\ddot{O}T}(s)=X(s) \ed \bd M_{d}^{F\ddot{O}T}=N_{i}(s) \ed
\bd M_{n}^{F\ddot{O}T}(s)=M(s) \ed \bd
N_{o}^{F\ddot{O}T}(s)=N_{o}^{-1}(s),
\ed
and the optimal controller, $C$, for the two block problem (\ref{twobl}) is
the inverse of optimal controller for the dual problem in \cite{FOT}, i.e.,
$\left(C_{opt}^{F\ddot{O}T}\right)^{-1}$.

If we only consider the one block problem case, with $W_2=0$, then the
minimization of
\bd \|W_{1}(Y-N_{i}Q_{1})\|_{\infty}
\ed
is simply a finite dimensional problem. On the other hand,  minimizing
\bd
\| W_{2}(1-M(Y-N_{i}Q_{1})) \|_\infty
\ed
is an infinite dimensional problem.

\section{Example}

In this section, we illustrate the computation of $\Hi$ controllers for
systems with infinitely many right half plane poles. The example is a
plant containing an internal delayed feedback:
\bd
P(s)=\frac{R(s)}{1+e^{-hs}R(s)}
\ed
where $R(s)=k\left(\frac{s-a}{s+b}\right)$ with $k>1$, $a>b>0$ and
$h>0$. Note that the denominator term $(1+e^{-hs}R(s))$ has infinitely
many zeros $\sigma_n \pm j \omega_n$, where $\sigma_n \rightarrow
\sigma_o= \frac{\ln(k)}{h}>0$, and $\omega_n \rightarrow (2n+1)\pi$, as
$n\rightarrow \infty$.  Clearly, $P(s)$ has only one right half plane
zero at $s=a$.

The plant can be written as explained in Section~2,
\be
\label{eq:Ps}
P(s)=\frac{N_i(s)}{M(s)}N_o(s)
\ee
where
\bea
\nonumber N_i(s)&=&\left(\frac{s-a}{s+a}\right) \\
\nonumber N_o(s)&=&\frac{1}{1+{(s-b)\over k(s+a)}e^{-hs}} \\
\nonumber M(s)&=&\frac{(s+b) + k(s-a)e^{-hs}}{(s-b)e^{-hs}+k(s+a)}
\eea
It is clear that $N_o$ is invertible in $H^\infty$, because
$\|\frac{s-b}{k(s+a)}\|_\infty <1$. By the same argument, $M$ is
stable. To see that $M$ is inner,  we write it as
\bd
M(s)={m(s)+f(s) \over 1+ m(s) f(-s)}
\ed
with $m(s)=\left({s-a\over s+a}\right)e^{-hs}$, and $f(s)={s+b\over
k(s+a)}$. Note that $m(s)$ is inner, $m(s)f(-s)$ is stable, and
$M(s)M(-s)=1$. Thus $M$ is inner, and it has infinitely many zeros in
the right half plane.

The optimal $\Hi$ controller can be designed for weighted sensitivity
minimization problem in (\ref{eq:wsm}) where $P$ is defined in
(\ref{eq:Ps}) and weight functions are chosen as $W_1(s)=\rho$,
$\rho>0$ and $W_2(s)=\frac{1+\alpha s}{\beta+s}$, $\alpha>0$,
$\beta>0$, $\alpha \beta<1$. As explained before, this problem can be
solved by the method in \cite{FOT} after necessary assignments are
done, $ W_{1}^{F\ddot{O}T}(s)=\frac{1+\alpha s}{\beta+s}$,
$W_{2}^{F\ddot{O}T}(s)=\rho$, $M_{d}^{F\ddot{O}T}=\frac{s-a}{s+a}$,
\bd
M_{n}^{F\ddot{O}T}(s)=\frac{(s+b)+k(s-a)e^{-hs}}{(s-b)e^{-hs}+k(s+a)}
\ed
\bd
N_{o}^{F\ddot{O}T}(s)=\frac{(s-b)e^{-hs}+k(s+a)}{k(s+a)}.
\ed

We will briefly outline the procedure to find the optimal $\Hi$
controller.

\begin{description}
\item [1)] Define the functions,
\bd F_\gamma(s)= \gamma
\left(\frac{\beta-s}{a_{\gamma}+b_{\gamma} s}\right), ~~~~~~
\omega_\gamma = \sqrt{\frac{1-\gamma^2
\beta^2}{\gamma^2-\alpha^2}}~~~~{\rm for}~~\gamma>0
\ed
where $a_{\gamma}=\sqrt{1+\rho^2 \beta^2-\rho^2 \gamma^{-2}}$, ~and~~
$b_{\gamma}=\sqrt{(1-\rho^2 \gamma^{-2})\alpha^2+\rho^2}$.
\item [2)] Calculate the minimum singular value of the matrix,
\bd \footnotesize{M_\gamma=\left[ \begin{array}{cccc}
  1 & j \omega_{\gamma} & M(j \omega_{\gamma})F_\gamma(j \omega_{\gamma})
  & j \omega_{\gamma} M(j \omega_{\gamma})F_\gamma(j \omega_{\gamma}) \\
  1 & a & M(a)F_\gamma(a) & a M(a)F_\gamma(a) \\
  M(j \omega_{\gamma})F_\gamma(j \omega_{\gamma})
  & -j \omega_{\gamma} M(j \omega_{\gamma})F_\gamma(j \omega_{\gamma})
  & 1 & -j \omega_{\gamma} \\
  M(a)F_\gamma(a) & -a M(a)F_\gamma(a) & 1 & -a
\end{array} \right] }
\ed for all values of $\gamma\in (
\max\{\alpha,\frac{\rho}{\sqrt{1+\rho^2
\beta^2}}\},\frac{1}{\beta})$. The optimal gamma value,
$\gamma_{opt}$, is the largest gamma which makes the matrix
$M_\gamma$ singular.
\item [3)] Find the eigenvector
$l=[l_{10}, l_{11}, l_{20}, l_{21}]^T$ such that
$M_{\gamma_{opt}}l=0$. 
\item [4)] The optimal $\Hi$ controller can be written as,
\bd
C_{opt}(s) =
\frac{k_f+K_{2,FIR}(s)}{K_1(s)}
\ed
where $k_f$ is constant, $K_1(s)$ is finite dimensional, and
$K_{2,FIR}(s)$ is a filter whose impulse response is of finite duration
\bea
\nonumber K_1(s)&=&\frac{k(l_{21}s+l_{20})}{\gamma_{opt}(\beta+s)}, \\
\nonumber k_f &=&
\left( \frac{k b_{\gamma_{opt}} l_{11}-\gamma_{opt}l_{21}}
                                    {\gamma_{opt}^2-\alpha^2}\right), \\
\nonumber K_{2,FIR}(s)&=&A(s)+B(s)e^{-hs}, \\
\nonumber k_f+A(s)&=&
\frac{k(s+a)(a_{\gamma_{opt}}+b_{\gamma_{opt}}s)(l_{11}s+l_{10})
                                +\gamma_{opt}(\beta-s)(l_{21}s+l_{20})(s+b)}
{((1-\gamma_{opt}^2\beta^2)+(\gamma_{opt}^2-\alpha^2)s^2)(s-a)}, \\
\nonumber B(s)&=&
\frac{(s-b)(a_{\gamma_{opt}}+b_{\gamma_{opt}}s)(l_{11}s+l_{10})
                               +k\gamma_{opt}(\beta-s)(l_{21}s+l_{20})(s-a)}
{((1-\gamma_{opt}^2\beta^2)+(\gamma_{opt}^2-\alpha^2)s^2)(s-a)}.
\eea
\end{description}

As a numerical example, if we choose the plant as
$P(s)=\frac{2\left(\frac{s-3}{s+1}\right)}
{1+2\left(\frac{s-3}{s+1}\right)e^{-0.5s}}$ and the weight functions
as $W_1(s)=0.5$, $W_2(s)=\frac{1+0.1s}{0.4+s}$, then the optimal $\Hi$
cost is $\gamma_{opt}=0.5584$, and the corresponding controller is
\bd
C_{opt}(s)=\left(\frac{0.558s+0.223}{2s+3.725}\right)\left(1.477+K_{2,FIR}(s) \right)
\ed
where 
\bd
K_{2,FIR}(s)=\frac{(2.0807s^2-6.3022s-0.8264)
                    -(0.6147s^3-0.7682s^2-5.2693s+1.5870)e^{-0.5s}}
             {(0.3018s^3-0.9053s^2+0.9501s-2.8504)}.
\ed
whose impulse response is of finite duration:
\bd
\mathcal{L}^{-1}(K_{2,FIR}(s))=
   \left\{ \begin{array}{cc}
  -0.27e^{3t}+7.16\cos(1.77t) + 0.36\sin(1.77t)-2.037\delta(t-0.5)& 0\leq t \le 0.5\\
   0 & t>0.5
\end{array} \right. .
\ed

\section{Conclusions}

In this note we have considered $\Hi$ control of a class of systems
with infinitely many right half plane poles. We have demonstrated that
the problem can be solved by using the existing $\Hi$ control
techniques for infinite dimensional systems with finitely many right
half plane poles. An example from delay systems is given to illustrate
the computational technique.

\bigskip
\noindent {\bf Acknowledgement}: The problem discussed in this
note was posed by Professor Kirsten Morris via a private
communication. We also thank an anonymous reviewer for checking
the numerical example carefully, and pointing out a missing
argument in the original version of the paper.

\end{document}